\documentclass[aps,prb,reprint,superscriptaddress,showpacs,floatfix]{revtex4-1}  
\usepackage{amsmath,amssymb,xcolor}
\usepackage[colorlinks,bookmarks=false,citecolor=red,linkcolor=blue,urlcolor=blue]{hyperref}
\usepackage{graphicx,epstopdf}
\usepackage{hyperref}
\usepackage{ulem}
\hypersetup{
    colorlinks,
    citecolor=blue,
    filecolor=blue,
    linkcolor=blue,
    urlcolor=blue} 
\usepackage{array,multirow}
\newcolumntype{C}[1]{>{\centering\arraybackslash}m{#1}}
\begin{document}
\title{Singlet quantum phases and magnetization of the frustrated spin-1/2 ladder with ferromagnetic (F) exchange in legs and alternating F-AF exchange in rungs}

\author{Monalisa Chatterjee}
\affiliation{S. N. Bose National Centre for Basic Sciences, Block-JD, Sector-III, Salt Lake, Kolkata 700106, India}

\author{Manoranjan Kumar}
\email{manoranjan.kumar@bose.res.in}
\affiliation{S. N. Bose National Centre for Basic Sciences, Block-JD, Sector-III, Salt Lake, Kolkata 700106, India}

\author{Zolt\'an G. Soos}
\email{soos@princeton.edu}
\affiliation{Department of Chemistry, Princeton University, Princeton, New Jersey 08544, USA}

\begin{abstract}
The magnetization $M(h)$ is used to identify three singlet quantum phases of the ladder with isotropic exchange interactions. The Dimer phase with frustrated F exchanges in rungs and legs has a first-order $M(h)$ transition at $0$ K from singlet to ferromagnetic at the saturation field $h_s$. The Haldane-DAF phase with strong F exchange in rungs and net AF exchange between rungs has continuous $M(h)$ and is adiabatically connected to the $S = 1$ Heisenberg AF chain. The AF phase with strong F exchange in legs and net AF exchange between legs has continuous $M(h)$ and is adiabatically connected to the spin-1/2 $J_1-J_2$ model with $J_1 > 0$ and $J_2 < 0$. All three singlet phases have finite gaps to the lowest triplet state. 
\end{abstract}
\maketitle
\section{Introduction}

The spin-1/2 Heisenberg antiferromagnetic (AF) chain with isotropic exchange $J > 0$ between first neighbors has been central to theoretical studies of correlated many-spin systems, including famous exact solutions, and the magnetism of inorganic and organic materials that contain 1D spin-1/2 chains. The scope of many-spin research has vastly grown in recent decades to include, among many others: ladders and 2D systems, competing exchanges of either sign, spin $S = 1$ or greater, topology and frustration, new theoretical and numerical methods. Quantum phases and phase diagrams have been a major thrust. Low dimensional spin systems have exotic phases both in zero and applied magnetic field.

\begin{figure}[h!]
\includegraphics[width=\linewidth]{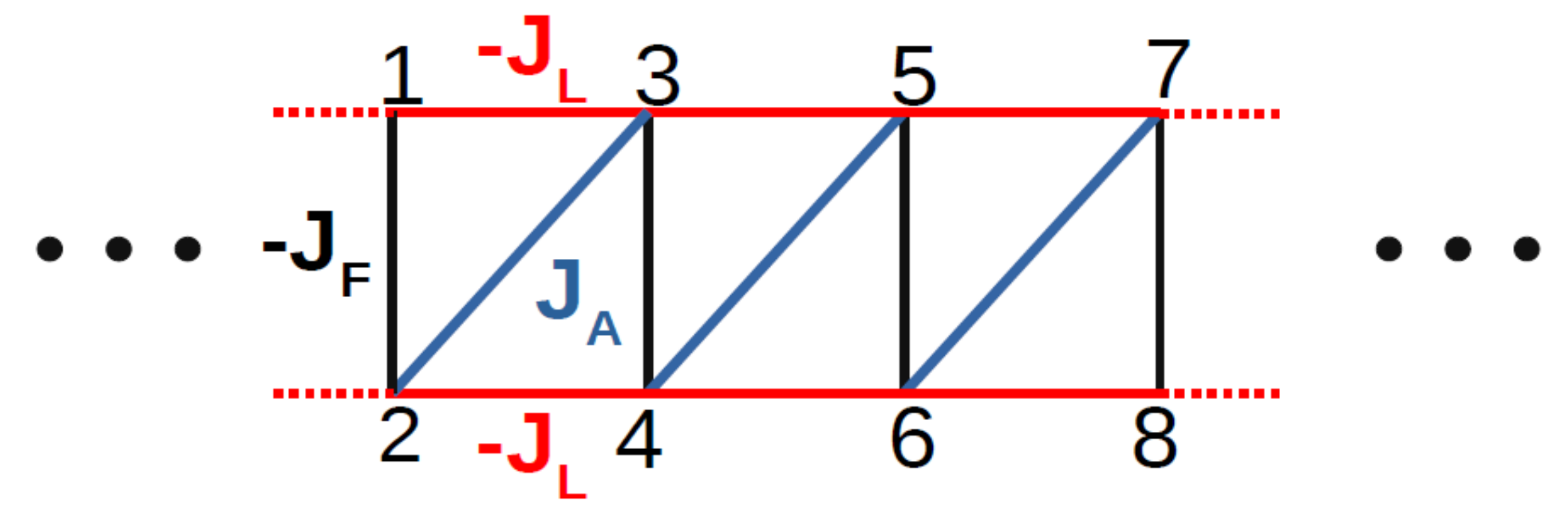}
\caption{The F-AF spin-1/2 ladder with F exchange $-J_L < 0$ between spins $r$ and $r + 2$ in either leg, F exchange $-J_F$ in rungs $2r - 1$, $2r$ and AF exchange $J_A$ in rungs $2r$, $2r + 1$.
}
\label{ssm_fig}
\end{figure}

In this paper we discuss the singlet phases of the frustrated F-AF ladder \cite{Dmitriev2000, Hida2013} in Fig. \ref{ssm_fig}. The ladder has two noteworthy features: the ground state is either a singlet or ferromagnetic and the exact ground state is a product of singlet-paired spins along a line where F exchanges cancel exactly. Both features are central in the following. A product of singlet-paired spins is the exact ground state \cite{ Majumder1969,Chitra1995,SRIRAMSHASTRY1981,Furukuwa2012,Sahoo2020} at special points of other 1D and 2D spin systems.
The ladder has $2N$ spins-1/2 with isotropic ferromagnetic (F) exchange $-J_L$ between neighbors in legs, F exchange $-J_F$ at rungs $2r -1$, $2r$ and AF exchange $J_A$ at rungs $2r$, $2r + 1$. We set $J_A = 1$ as the unit of energy and impose periodic boundary conditions,
\begin{eqnarray}
\label{ssm_eq1}
{H_{F-AF}(J_L,J_F)}=\sum_{r=1}^{N} (\vec{S}_{2r} \cdot \vec{S}_{2r+1}-J_F \vec{S}_{2r-1}  \cdot \vec{S}_{2r})+ \nonumber \\ 
& & \hspace*{-160pt} \sum_{r=1}^{2N} (-J_L \vec{S}_r \cdot \vec{S}_{r+2}-hS_r^z).
\end{eqnarray}
The ladder is equivalent to a $1D$ chain with two spins per unit cell, exchange $-J_L$ between second neighbors and alternating exchanges $-J_F$ and $J_A$ between first neighbors. The total spin $S \leq N$ and its z component $S^z$ are conserved. The last term is the interaction with a magnetic field $h$.

The exchanges $J_A = 1$, $J_L$ and $J_F$ dominate in different regions of the positive quadrant of the $J_L$, $J_F$ plane. Near the origin, the spins $2r$, $2r + 1$ are singlet paired with ground state energy $\epsilon_0(J_L,J_F)$ per dimer and singlet-triplet gap $\epsilon_T(J_L,J_F)$;
$\epsilon_0(J_L,J_F) = \epsilon_0(0,0) = -3/4$ is exact along $J_L = J_F/2 \leq 1$. F exchanges cooperate in the F phase with aligned spins and energy per dimer
\begin{equation}
{\epsilon_F(J_L,J_F)}= -(2J_L+J_F-1)/4.
\label{ssm_eq2}
\end{equation}
It follows that large $J_L$ cannot induce the F phase if the net exchange $1 - J_F > 0$ between spins in different legs is AF; likewise, large $J_F$ cannot induce the F phase if the net exchange $1 - 2J_L > 0$ between rungs is AF.

The relation $\epsilon_0(J_L,J_F) = \epsilon_F(J_L,J_F)$ defines the singlet/F phase boundary. Dmitriev et al. obtained the exact boundary and showed that intermediate $S$ has higher energy. \cite{Dmitriev2000} In the present notation
\begin{equation}
{J_F}= 2J_L/(2J_L-1),    \hspace{0.8cm} 2J_L\geq 1.
\label{ssm_eq3}
\end{equation}       
Eq. \ref{ssm_eq3} is also the nonmagnetic/F boundary of classical spins. \cite{Hida2013}

We find below that the ladder has three singlet quantum phases: AF, Dimer and Haldane-DAF. All three have nondegenerate ground states and finite $\epsilon_T(J_L,J_F)$. We distinguish among them using the magnetization $M(h)$ per dimer, not considered previously for the ladder, to fully aligned spins at the saturation field $h_s$. The $T = 0$ K magnetization transition of the Dimer phase is first order, discontinuous at $M(h_s)$, while the Haldane-DAF and AF phases have continuous $M(h)$.

The quantum phase diagram is presented in Sec. \ref{sec-II} together with that of classical spins. The $M(h)$ analysis in Sec.  \ref{sec-III}  is based on exact solution of Eq. \ref{ssm_eq1} with periodic boundary conditions up to $2N = 24$ spins and density matrix renormalization group (DMRG) calculations on longer ladders. We consider spin correlations in Sec. \ref{sec-IV} and rigorously relate the Haldane-DAF phase to the spin-1 Heisenberg AF chain and the AF phase to the $J_1-J_2$ model. Sec. \ref{sec-V} is a brief summary.

\section{Quantum phase diagram \label{sec-II}}
\begin{figure}[h!]
\includegraphics[width=\linewidth]{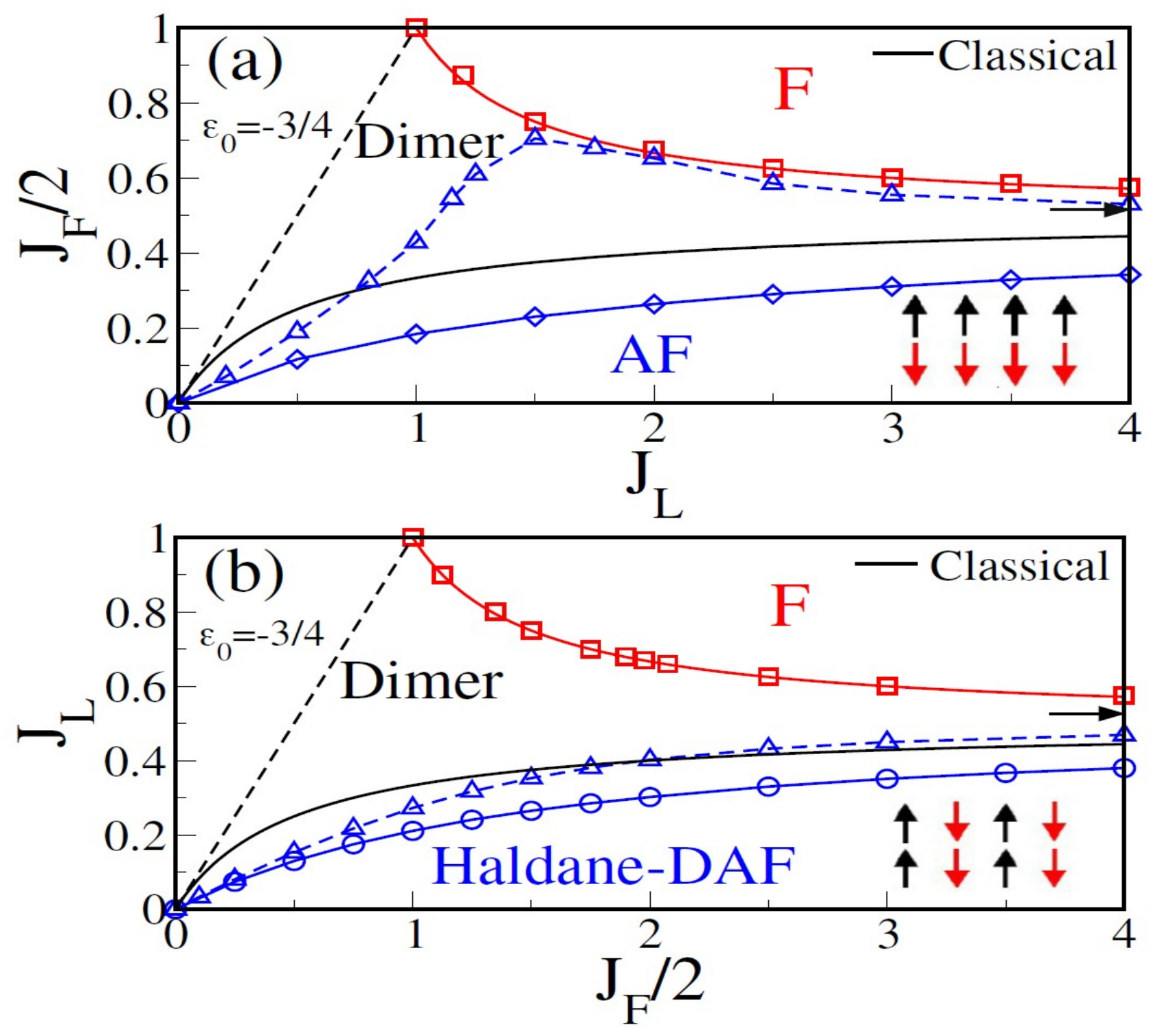}
\caption{Quantum phase diagram of the ladder for (a) $J_L \geq J_F/2$ and (b) $J_F/2 \geq J_L$. The red line is the singlet/F boundary, Eq. \ref{ssm_eq3}. Solid black lines are the AF/Spiral and DAF/Spiral boundaries of classical spins. \cite {Hida2013} The exact  $\epsilon_0 = -3/4$ is along the dashed black line. Solid blue lines are Eq. \ref{ssm_eq6} for the AF/Dimer boundary in (a) and the Haldane-DAF/Dimer boundary in (b). The arrows mark $J_F/2 = J_L = 1/2$. $M(h)$ is discontinuous in the Dimer phase. It is a step function from the dashed blue lines, Eq. \ref{ssm_eq5}, to the F boundary.}
\label{ssm_phase_diagram}
\end{figure}

The singlet phases of Eq. \ref{ssm_eq1} in the $J_L$, $J_F$ plane are necessarily close to the origin, where $J_L$ and $J_F$ are comparably small, or close to the $J_L$ or $J_F$ axis, where the other exchanges are small. To discuss ranges of $J_L$ and $J_F$, we show the phase diagram for $J_L \geq J_F/2$ in Fig. \ref {ssm_phase_diagram} (a) and for $J_F/2 \geq J_L$ in Fig. \ref {ssm_phase_diagram} (b). The red line is the singlet/F boundary, Eq. \ref{ssm_eq3}, that is also the nonmagnetic/F boundary of classical spins; the calculated points are exact for a ladder with 24 spins. The dashed black line is $J_L = J_F/2 \leq 1$ with $\epsilon_0 = -3/4$.  

It is instructive to contrast the singlet phases of quantum spins with the exact nonmagnetic phases of classical spins, \cite {Hida2013} which are shown as solid black lines. As sketched in Fig. \ref {ssm_phase_diagram} (a), classical spins in the AF phase are antiparallel in legs and correspond to the N\'eel state with spins ...$\alpha\beta\alpha\beta$..., although quantum fluctuations preclude long range order in $1D$ quantum systems. The DAF phase of classical spins in Fig. \ref {ssm_phase_diagram} (b) has parallel spins in rungs $2r - 1$, $2r$ and antiparallel rungs that correspond to ...$\alpha\alpha\beta\beta$... along the ladder. Aligned spins at $2r - 1$ and $2r$ suggest a spin-1 chain with AF exchange whose singlet-triplet gap was predicted by Haldane to be finite. \cite{Haldane1983}   

The solid blue line in Fig. \ref {ssm_phase_diagram} (a) is the Dimer/AF phase boundary. The magnetization $M(h)$ increases up to the saturation field $h_s$ at which the absolute ground state becomes the lowest Zeeman component of the F state. In the thermodynamic limit, $M(h)$ is continuous at $h_s$ from $J_L = 0$ to the AF/Dimer boundary where it becomes discontinuous. The discontinuity 
increases up to the dashed blue line in Fig. \ref {ssm_phase_diagram} (a) where M(h) becomes a step function at $h_s$. The saturation field decreases to $h_s= 0$ at the Dimer/F boundary since an infinitesimal field induces the F phase.
 
The Haldane-DAF phase in Fig. \ref {ssm_phase_diagram} (b) has continuous $M(h)$ at $h_s$ from $J_F = 0$ to the Dimer boundary shown as a solid blue line. The Dimer phase has discontinuous $M(h)$ from the solid blue line to the F boundary; it is a step function from the dashed blue line to the boundary.

We emphasize the different origins of the nonmagnetic phases of classical spins and the singlet phases of quantum spins. Eq. \ref{ssm_eq1} at $h = 0$ is minimized at $J_L$, $J_F$ with respect to the orientation of classical spins. \cite {Hida2013} The $0$ K magnetization $M(h)$ of Eq. \ref{ssm_eq1} at $J_L$, $J_F$ is obtained up to $h_s$ for quantum spins. Although similar in some respect, the singlet and nonmagnetic phases are fundamentally different. We speculate that spin 
fluctuations explain the reduced area of quantum phases in Fig. \ref {ssm_phase_diagram} compared 
to the AF and DAF phases of classical spins. For completeness, we note
that Hida et al. \cite {Hida2013} have discussed several topological phases of ladders with $J_F > 1/2, J_L > 1$. Topological phases are beyond the scope of the present study.

\section{Magnetization \label{sec-III}}
We use two numerical methods, ED and DMRG, to solve Eq. \ref{ssm_eq1} at $J_A = 1$ and variable $J_L$, $J_F$ in sectors with $0 \leq S = S^z \leq N$. Exact diagonalization at $2N = 16$ and $24$ spins returns the lowest energies $E(S^z,2N)$ and generates a preliminary phase diagram. DMRG is then used for larger systems.

DMRG is a well-established numerical technique \cite{White1992,Schollwock2005,Hallberg2006} for the ground state and low-lying excited states of correlated 1D systems. We use a modified DMRG algorithm \cite {Sudip2019} that adds four new sites (instead of two) to the superblock at each step. This avoids interaction terms between old blocks in models with second neighbor exchange. All calculations are performed with periodic boundary conditions. We obtain truncation errors of $10^{-10}$ on keeping 512 eigenvectors of the density matrix and 4 or 5 finite sweeps. Systems up to $2N = 64$ spins were used for finite-size scaling of phase boundaries.

In systems with isotropic exchange, the field dependence of the lowest Zeeman component of any state $E(S,2N)$ with spin $S$ and $S^z = S$ is 
\begin{equation}
{E(S^z,h,2N)}= E(S,2N)-hS^z,  \hspace{0.3cm} 0\leq S^z = S \leq N.
\label{ssm_eq4}
\end{equation}
We take $J_L$, $J_F$ leading to a singlet ground state at $h = 0$ and obtain the lowest energies $E(S^z,h,2N)$. Increasing $h$ generates multiple level crossings until $S = S^z = N$ becomes the absolute ground state.  

Since the field dependence of $E(N,h,2N)$ is extensive, the stabilization per dimer of the F state is
$\epsilon_F - h$. The field-induced crossing with the singlet ground state is at $h = \epsilon_F -
\epsilon_0$. On the other hand, $h = \epsilon_T$ is the field-induced crossing of the triplet and singlet ground states. At equal $h$ we obtain
\begin{equation}
{\epsilon_T(J_L,J_F)}= \epsilon_F(J_L,J_F)-\epsilon_0(J_L,J_F).
\label{ssm_eq5}
\end{equation}
If $\epsilon_F ­- \epsilon_0 < \epsilon_T$, $M(h)$ is a step function at $h_s$. Conversely, if 
$\epsilon_T$ is smaller, $M(h) = 1/N$ at $h = \epsilon_T$ and it is not a step function. The dashed blue lines in Fig. \ref {ssm_phase_diagram} are based on Eq. \ref{ssm_eq5}. $M(h)$ is a step function from the dashed lines to the Dimer/F boundary.

$M(h)$ is continuous at $h_s$ if the $S^z$ crossing between $N - 1$ and $N - 2$ occurs at lower
$h$ than the crossing between $N$ and $N - 1$ at $h_s$. If instead $h_s < h$, there is a 2/N discontinuity at $M(h_s)$ and $S^z = N - 1$ is never the ground state. The condition $h = h_s$ leads to
\begin{equation}
{E}(N,2N)+{E}(N-2,2N)= 2{E}(N-1,2N).
\label{ssm_eq6}
\end{equation}
$M(h_s)$ is discontinuous if $2E(N-1,2N)$ is larger than the left-hand side. Eq. \ref {ssm_eq6} generates the AF/Dimer and Haldane-DAF/Dimer phase boundaries in Fig. \ref{ssm_phase_diagram} (a) and (b), respectively. The thermodynamic limit of Eq. \ref {ssm_eq6} has a prominent role in strongly correlated spin chains. \cite{Parvej2016,Onishi2015,Parvej2017,Sato2011} Due to attractive interactions or binding energy, the double excitation $E(N­-2,2N)$ is lower than two single excitations when $2E(N-1,2N)$ is larger.

The saturation field is size independent. The Dimer phase in Fig. \ref{ssm_phase_diagram} has $h_s = 0$ at the Dimer/F boundary and finite $h_s$ elsewhere, with $h_s = 1 - J_L = 1 - J_F/2$ along the line $\epsilon_0= -3/4$. The AF phase in Fig. \ref{ssm_phase_diagram} (a) has $h_s = 1 - 2J_L \geq 0$ that offsets the net AF exchange between legs. The Haldane-DAF phase in Fig. 2 (b) has $h_s = 1 - J_F \geq 0$ that offsets the net AF exchange between rungs.

\begin{figure}[h]
\includegraphics[width=\linewidth]{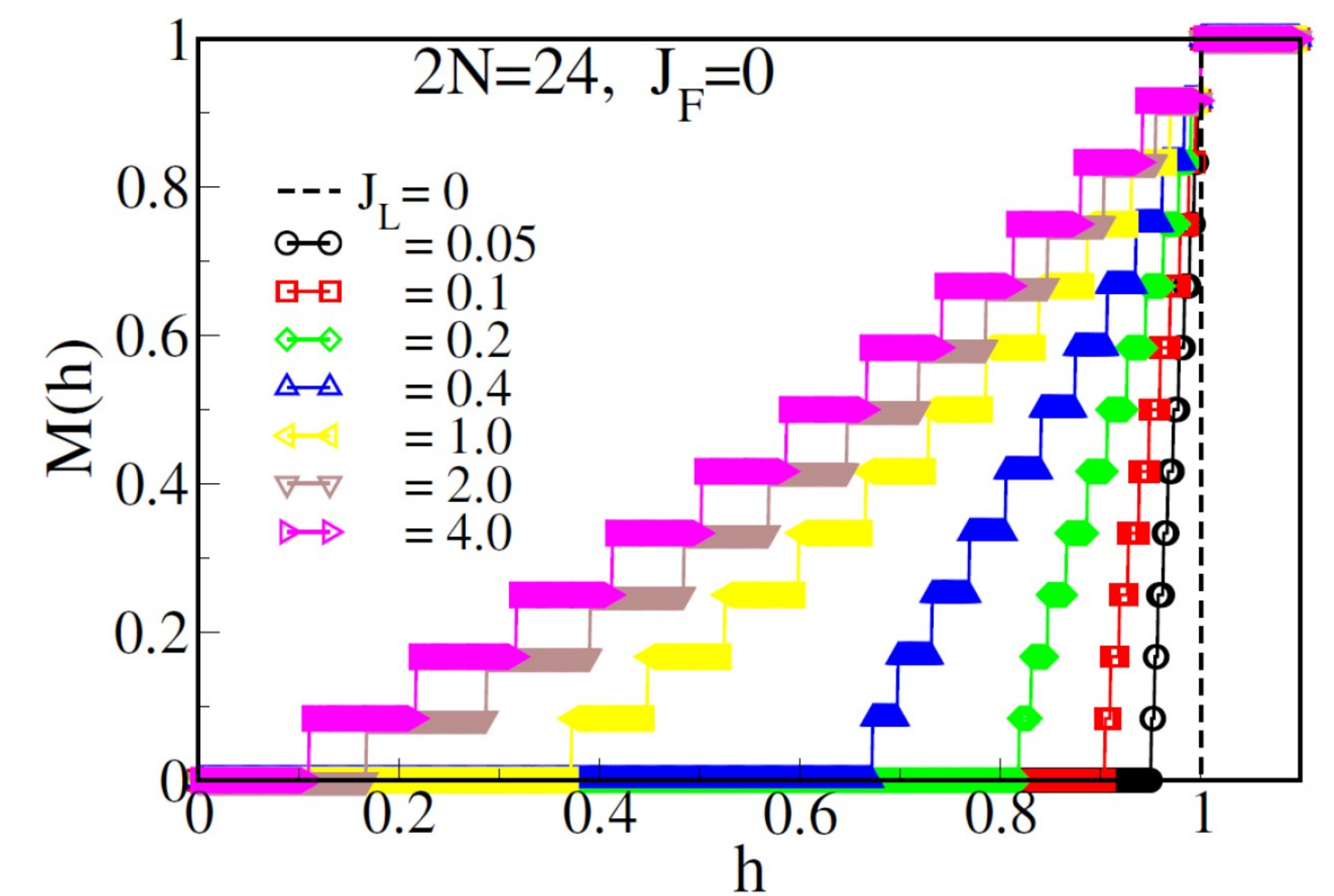}
\caption{Magnetization $M(h)$ of Eq. \ref{ssm_eq1} with $2N = 24$ spins, $J_F = 0$ and variable $J_L \geq 0$. $M(h)$ is a step function at $h_s = 1$ when $J_L = 0$ and continuous when $J_L > 0$.
}
\label{mag_jf_0}
\end{figure}

The magnetization curves in Fig. \ref{mag_jf_0}  are for $J_F = 0$, $2N = 24$ spins in Eq. \ref{ssm_eq1} and variable $J_L \geq 0$.  The $N$ steps of $M(h)$ correspond to reversing 12 spins. Increasing the system size increases the number of steps but not the saturation field $h_s = 1$, a step function at $J_L = 0$. Exchange $J_L > 0$ between dimers generates a triplet band with $\epsilon_T(J_L,0) < 1$ and continuous $M(h)$ in the thermodynamic limit. Increasing $J_L$ lowers the $M(h) > 0$ threshold. Continuous $M(h)$ with $h_s = 1$ are also found at $J_L = 0$ and variable $J_F > 0$. 

\begin{figure}[h]
\includegraphics[width=\linewidth]{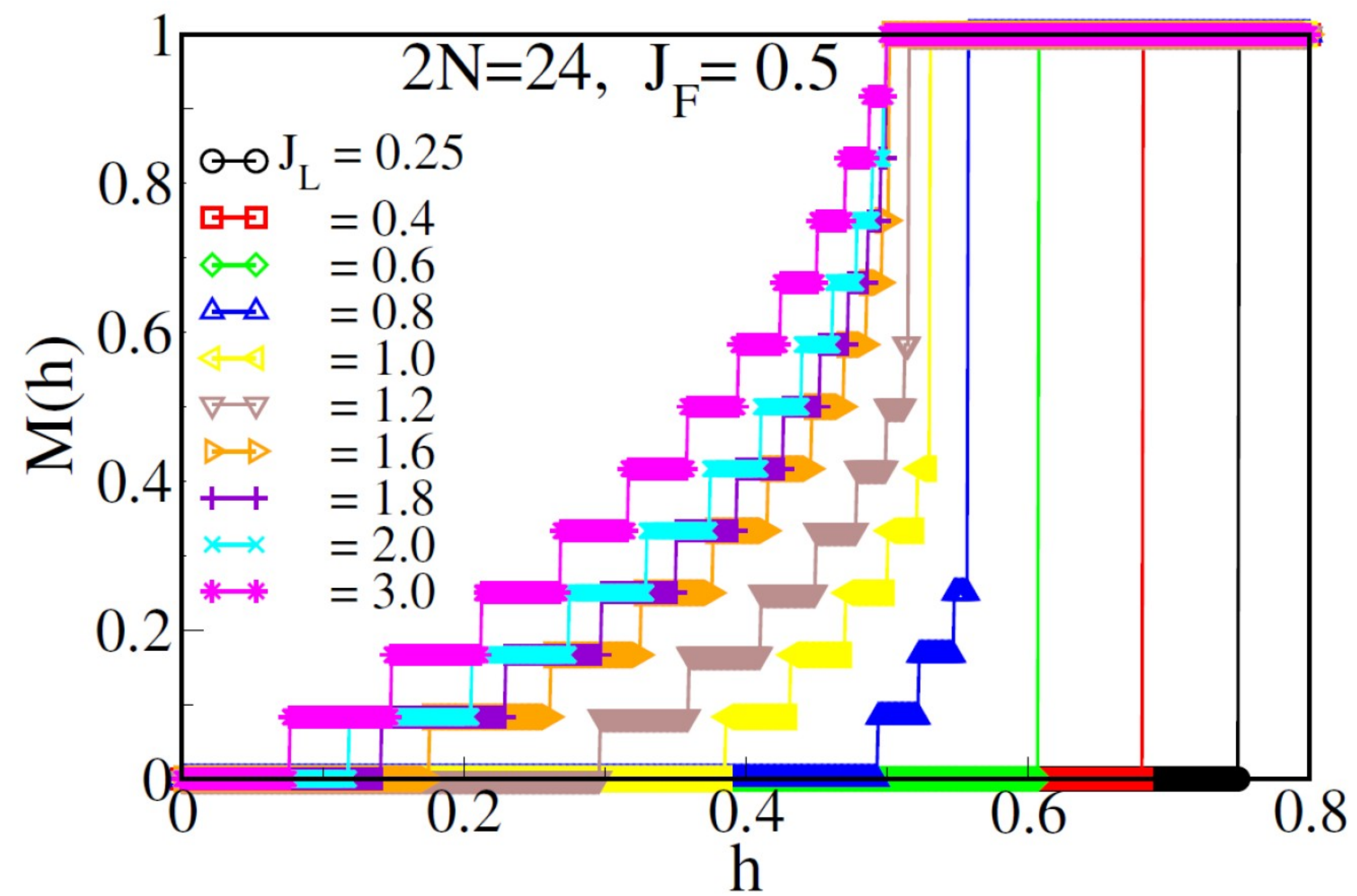}
\caption{Magnetization $M(h)$ of Eq. \ref{ssm_eq1} with $2N = 24$ spins, $J_F = 1/2$ and variable $J_L \geq J_F/2$. $M(h)$ is a step function from $h_s = 0.75$ at $J_L = 0.25$ to $h_s = 0.59$ at $J_L = 0.640$. $M(h$) is discontinuous at $h_s = 0.5$ up to $J_L = 1.81$.}
\label{mag_jf_05}
\end{figure}

Fig. \ref {mag_jf_05}  illustrates discontinuous $M(h)$ in the Dimer phase and the Dimer/AF boundary, specifically for $J_F = 1/2$, $2N = 24$ spins and variable $J_L \geq J_F/2 $. $M(h)$ is a step function at $h_s = 3/4$ when $J_L = 1/4$ and $\epsilon_0(1/4,1/2) = -3/4$. Increasing $J_L$ lowers saturation field to $h_s = 1 - J_F = 1/2$. $M(h)$ is a step function until $\epsilon_T = \epsilon_F-\epsilon_0 $ in Eq. \ref{ssm_eq5}, at $J_L = 0.640$ for $2N = 24$, already in the thermodynamic limit. At larger $J_F = 1.5$, we find $J_L = 1.318$ at $2N = 24$, $1.395$ at $2N = 64$ and $1.41$ in the limit. Points $(J_L,J_F)$ in the limit generate the dashed blue lines in Fig. \ref {ssm_phase_diagram} (a) for $J_L \geq J_F/2$. The Dimer/AF boundary at which $M(h)$ becomes continuous is given by Eq. \ref{ssm_eq6}, at $J_L = 1.81$ for $2N = 24$, also in the limit.  

The same procedure holds for variable $J_F/2 \geq J_L$ at constant $J_L$ in Fig. \ref {ssm_phase_diagram} (b). $M(h)$ is initially a step function at $h_s = 1 - J_L$ when $J_L > 1/2$
and remains a step function with increasing $J_F$ up to the Dimer/F boundary. The saturation field decreases with increasing $J_F$ to $h_s = 1 - 2J_L$ when $J_L \leq 1/2$. $M(h)$ is discontinuous but not a step function when $\epsilon_T = \epsilon_F-\epsilon_0$, Eq. \ref{ssm_eq5}, and it remains discontinuous with increasing $J_F$ until the Dimer/Haldane-DAF boundary, Eq. \ref{ssm_eq6}, is reached. Convergence to the thermodynamic limit is faster for weak AF exchange between F rungs in Fig. \ref {ssm_phase_diagram} (b) than for weak AF between F legs in Fig. \ref {ssm_phase_diagram} (a).

\section{ Haldane-DAF and AF phases  \label{sec-IV}}
The singlet quantum phases of the F-AF ladder are based on $M(h)$. The Haldane-DAF phase extends to arbitrarily large $J_F$ provided that $2J_L < 1$; the AF phase to large $J_L$ when $J_F < 1$. We study the limits using $ \epsilon_T(J_L,J_F)$ and spin correlations functions $C(m,n)$
\begin{equation}
{C(m,n)}=\langle \vec {S}_m  \cdot \vec {S}_n \rangle.
\label{ssm_eq7}
\end{equation}
The expectation value is with respect to the singlet ground state of Eq. \ref{ssm_eq1} at $J_L$, $J_F$. The ladder with periodic boundary conditions has three bond orders: $C(2,3)$ for rungs $2r$, $2r + 1$ in Fig. \ref {ssm_fig}, $C(1,2)$ for rungs $2r - 1$, $2r$ and $C(1,3)$ for $r$, $r + 2$ in either leg. 

All $C(m,n)$ are zero at $J_L = J_F/2 \leq 1$ except $C(2,3) = -3/4$. In Fig. \ref {ssm_phase_diagram} (a) with $J_L > J_F/2$, the $C(m,n$) are $\geq 0$ $( \leq 0)$ when $n,m$ are in
the same (different) leg. When $J_F > 2J_L$, the bond orders are $C(1,2) > 0$ and $C(2,3)$, $C(1,3) < 0$. The sign of $C(m,n)$ for more distant spins follows the DAF pattern in Fig. \ref {ssm_phase_diagram} (b). The gap 
$\epsilon_T(J_L,J_F)$ is $1$ at $J_L = J_F = 0$ and decreases with either $J_L$ or $J_F$. The
$C(m,n)$ are very short ranged near the origin due to large $\epsilon_T$.

We consider large $J_F$ with triplet paired spins in rungs $2r - 1$, $2r$. The $h = 0$ ground state is $3^N$-fold degenerate at $J_A = J_L = 0$. The spin-1 Heisenberg chain has AF exchange $J$ between adjacent $S = 1$ rungs with Haldane gap \cite{White1993} $\Delta(1) = 0.4105$. Connections between ladders and the $S = 1$ chain have already been noted. \cite{Hida1992,Hida2013,Sahoo2020} Here we follow the quantitative evolution of the Haldane-DAF phase of the ladder to $S = 1$ chains at infinite $J_F$.

We rewrite Eq. \ref{ssm_eq1} at $h = 0$ as
\begin{eqnarray}
\label{ssm_eq8}
{H_{F-AF}(J_L,J_F)}= -J_F\sum_{r=1}^{N}  \vec{S}_{2r-1} \cdot \vec{S}_{2r}+ (1-2J_L)/4 \nonumber \\ 
&&\hspace*{-205pt} \times \sum_{r=1}^{N} (\vec{S}_{2r-1}+\vec{S}_{2r})\cdot (\vec{S}_{2r+1}+
\vec{S}_{2r+2})+V'.
\end{eqnarray}
The first term corresponds to noninteracting dimers. The second term is the $S = 1$ chain with AF exchange $(1 - 2J_L)/4 > 0$ between neighbors. The operator $V’$ contains all other exchanges. The coefficients are: $(3 + 2J_L)/4$ for exchange $2r$, $2r + 1$; $- (1 + 2J_L)/4$ for exchange $r$, $r + 2$; and $- (1 - 2J_L)/4$ for exchange $2r -1$, $2r + 3$. Since all $V'$ terms reverse a spin in each of two adjacent rungs, they lead in lowest order to an effective S = 1 Hamiltonian with excitations of order $1/J_F$. Eq. \ref{ssm_eq8} adiabatically connects ladders with finite $J_F$ and $2J_L < 1$ to $S = 1$ AF chains with $V’ = 0$ in the limit $J_F \to \infty$.

There are two limits: $J_F \to \infty$ at any system size and $N \to \infty$. We followed the $J_F$ dependence of $\epsilon_T(J_L,J_F)$ and of correlations between adjacent rungs at $2N = 16$. The ED results in Table \ref{tab:table1} are for $J_L = 0$ and $0.25$. The gaps and correlations converge smoothly, with $J_F = 200$ at the limit (for $2N = 16$). They match the correlations of the finite $S = 1$ chain, $-1.4171$, and gap $\Delta = 0.593555$. Spin correlations depend weakly on $J_L$ at finite $J_F$. The $J_L = 0$ and $1/4$ gaps in Eq. \ref{ssm_eq8} are $\Delta/4$ and $\Delta/8$, respectively, in quantitative agreement with the finite $S = 1$ chain. Identical $J_F \to \infty$ limits hold at different systems sizes and hence in the thermodynamic limit.

\begin{table}
\normalsize
\caption{\label{tab:table1} Spin correlations $\langle (S_1 + S_2).(S_3 + S_4)\rangle$ and gaps $\epsilon_T(J_L,J_F)$ of the ladder, Eq.\ref{ssm_eq1} with the $2N = 16$ spins, variable $J_F$ and $J_L = 0$ or $0.25$.}
\begin{center}
\begin{tabular}{|C{1.2cm}|C{1.6cm}|C{1.6cm}|C{1.6cm}|C{1.6cm}|}
\hline
&  \multicolumn {2}{|c|}{$\langle (S_1+S_2)\cdot (S_3+S_4) \rangle$} & \multicolumn {2}{|c|}{$\epsilon_T(J_L,J_F)$} \\
\hline
\hline
$J_F$ & $J_L=0$ & $J_L=0.25$ & $J_L=0$ & $J_L=0.25$  \\ 
\hline
$10$  & $-1.404$ & $-1.393$  & $0.187$ & $0.121$ \\
\hline
$40$  & $-1.416$ & $-1.416$  & $0.157$ & $0.083$ \\
\hline
$100$  & $-1.417$ & $-1.417$  & $0.152$ & $0.078$ \\
\hline
$200$  & $-1.417$ & $-1.417$  & $0.150$ & $0.076$ \\
\hline
$\infty$ & $-1.417$ &  $-1.417$ & $0.148$ & $0.074$ \\
\hline
$S=1$ & \multicolumn {2}{|c|}{$-1.417$} & $0.148$ & $0.074$ \\
\hline
\end{tabular}
\end{center}
\end {table}

The $J_F \to \infty$ limit equalizes the bond orders $C(2,3) = (1,3) = C(2,4) = C(1,4)$ among the constituent spins of adjacent $S = 1$ rungs. The limit enforces higher symmetry: a reflection or $C_2$ rotation that interchanges the legs in Fig. \ref {ssm_fig}. Correlations $C(m,n)$ among more distant $S = 1$ rungs also become equal. The gap $\epsilon_T(J_L,J_F)$ decreases with increasing $J_F$ and $J_L < 1/2$ to a finite limit that is $(1 - 2J_L)\Delta(1)/4 > 0$, proportional to  $\Delta(1) = 0.4105$. The gap explains faster size convergence in the Haldane-DAF phase in Fig. \ref {ssm_phase_diagram} (b). The phase has continuous $M(h)$, as does the $S = 1$ AF chain.   
 
The F-AF ladder has alternating exchanges $J_A = 1$ and $-J_F$ between first neighbors. The average is $(1 - J_F)/2 > 0$ when $J_F < 1$. As shown in Table \ref{tab:table2} for $2N = 16$ spins, the bond orders $C(1,2)$ and $(2,3)$ become equal as $J_L \to \infty$. The limit is a $J_1-J_2$ model with $J_1 = (1 - J_F)/2 > 0$ for first neighbors and $J_2 = -J_L$ for second neighbors. The limit enforces higher translational symmetry, one spin per unit cell instead of two.

At $J_L = 200$ we find $C_{av} = -0.3140$ at $J_F = 0$ and $-0.3132$ at $J_F = 0.5$. The $2N = 16$ $J_1-J_2$ model returns $C(1,2) = -0.3140$ at $J_2 = -200, J_1 = 0.5$ and $-0.3132$ at $J_2 = -200, J_1 = 0.5$, in quantitative agreement with $C_{av}$. The $J_1 = 0.5$ and $0.25$ spin gaps are $0.1254$ and $0.0626$ instead of $0.1255$ and $0.0630$ in Table \ref{tab:table2}. In the present context, the $J_L \to \infty$ limit has been reached (for $2N = 16$) by $J_L = -J_2 = 200$.

\begin {table}[h]
\normalsize
\caption{\label{tab:table2} 
Bond orders and gap $\epsilon_T(J_L,J_F)$ of the ladder, Eq.\ref{ssm_eq1}, with the $2N = 16$ spins, variable $J_L$ and $J_F = 0$ or $0.5$. $C_{av}= [C(1,2) + C(2,3)]/2$ and   $C_{dif}= C(1,2) - C(2,3).$}
\begin{center}
\begin{tabular}{|C{1.3cm}|C{1.2cm}|C{1.7cm}|C{1.7cm}|C{1.7cm}|} 
\hline
$J_L$ & $J_F$ & $-C_{av}$ & $C_{dif}$ & $\epsilon_T$  \\ 
\hline
\hline
 \multirow{2}{*}{$10$} & $0$ & $0.3419$ & $0.0297$ & $0.1360$ \\
\cline{2-5}
 &  $0.5$ & $0.3263$ & $0.0434$ & $0.0731$ \\
\hline
\multirow{2}{*}{$50$} & $0$ & $0.3180$ & $0.0057$ & $0.1271$ \\
\cline{2-5}
 &  $0.5$ & $0.3153$ & $0.0085$ & $0.0645$ \\
\hline
\multirow{2}{*}{$200$} & $0$ & $0.3140$ & $0.0007$ & $0.1255$ \\
\cline{2-5}
&  $0.5$ & $0.3140$ & $0.0021$ & $0.0626$ \\
\hline
\end{tabular}
\end{center}
\end {table}

Quite generally, $S =1/2$ chains with short-range isotropic exchange and one spin per unit cell are either gapless with a nondegenerate ground state or gapped with a doubly degenerate ground state. \cite {Allen1997,Mattis1965} The phase diagram of the $J_1-J_2$ model has been discussed previously. \cite{nomura1992,soos2016,Parvej2016} The nondegenerate ground state at $J_1 > 0$ and $J_2 < 0$ indicates a gapless chain. Alternating exchanges $J_A = 1$ and $-J_F$ lead to gapped chains with two spins per unit cell. It follows that the AF phase is gapped for finite $J_L$ but gapless as $J_L \to \infty$.  

The AF and Haldane-DAF phases of the ladder are directly related to models with higher discrete symmetry, the $S = 1$ AF chain as $J_F \to \infty$ and the $J_1-J_2$ model as $J_L \to \infty$.  As seen in Tables \ref{tab:table1} and \ref{tab:table2}, the limits are already emerging at $J_F = J_L = 10$, and lower symmetry increases $\epsilon_T(J_L,J_F)$.

The Dimer phase is bounded by the other phases in Fig. \ref {ssm_phase_diagram} except in the limits $J_L \to \infty$ , $J_F = 1$ or $J_F \to \infty $, $2J_L = 1$  when there is no net AF interaction, respectively, between legs or rungs $2r - 1$, $2r$. Exact cancellation along $J_L = J_F/2 \leq 1$ in the ground state leads to a step function $M(h)$ from the dashed blue lines in Fig. \ref {ssm_phase_diagram} to the Dimer/F boundary.

\section{Summary and conclusion \label{sec-V}}

The $0$ K magnetization $M(h)$ and saturation field $h_s$ indicate that the F-AF ladder, Eq. \ref{ssm_eq1} at $h = 0$, has the three singlet quantum phases shown in Fig. \ref {ssm_phase_diagram}.

    (a) $M(h)$ is continuous at $h_s$ for $J_L$, $J_F$ in the Haldane-DAF phase. Increasing $J_F$ at constant $J_L$ leads to an $S = 1$ AF chain with $J = (1 - 2J_L)/4 > 0$ in the limit $J_F \to \infty $.

    (b) $M(h)$ is continuous at $h_s$ for $J_L$, $J_F$ in AF phase. Increasing $J_L$ at constant $J_F$ leads to a $J_1-J_2$ chain with $J_2 = - J_L$ and $J_1 = (1 - J_F)/2 > 0$ in the limit $J_L \to \infty$ .

    (c) $M(h)$ is discontinuous at $h_s$ for $J_L$, $J_F$ in the Dimer phase. $M(h)$ is a step function between the dashed blue lines in Fig. \ref{ssm_phase_diagram} and the Dimer/F boundary, Eq. \ref{ssm_eq3}. The discontinuity decreases to zero at the Dimer/AF and Dimer/Haldane-DAF boundaries.

The singlet phases have finite gaps $\epsilon_T(J_L,J_F)$ that decrease from $\epsilon_T(0,0) = 1$ and vanish only in the $J_L \to \infty $ of the AF phase. The Haldane-DAF gap decreases with $J_F$ and is proportional to the $S = 1$ gap in the limit. 

ED and DMRG are well suited for $M(h)$ calculations that require the thermodynamic limits of Eq. \ref{ssm_eq6} for the Dimer/Haldane-DAF or Dimer/AF phase boundary and of Eq. \ref{ssm_eq5} for a step function $M(h)$ in the Dimer phase. Numerical accuracy remains challenging in systems under study with small $\epsilon_T(J_L,J_F)$ and spin correlations of intermediate range. 

\section{Acknowledgments}
M.K. thanks SERB for financial support through Grant Sanction No. CRG/2020/000754. M.C. thanks DST-INSPIRE for financial support.
\bibliography{reference}

\begin{thebibliography}{22}%
\makeatletter
\providecommand \@ifxundefined [1]{%
 \@ifx{#1\undefined}
}%
\providecommand \@ifnum [1]{%
 \ifnum #1\expandafter \@firstoftwo
 \else \expandafter \@secondoftwo
 \fi
}%
\providecommand \@ifx [1]{%
 \ifx #1\expandafter \@firstoftwo
 \else \expandafter \@secondoftwo
 \fi
}%
\providecommand \natexlab [1]{#1}%
\providecommand \enquote  [1]{``#1''}%
\providecommand \bibnamefont  [1]{#1}%
\providecommand \bibfnamefont [1]{#1}%
\providecommand \citenamefont [1]{#1}%
\providecommand \href@noop [0]{\@secondoftwo}%
\providecommand \href [0]{\begingroup \@sanitize@url \@href}%
\providecommand \@href[1]{\@@startlink{#1}\@@href}%
\providecommand \@@href[1]{\endgroup#1\@@endlink}%
\providecommand \@sanitize@url [0]{\catcode `\\12\catcode `\$12\catcode
  `\&12\catcode `\#12\catcode `\^12\catcode `\_12\catcode `\%12\relax}%
\providecommand \@@startlink[1]{}%
\providecommand \@@endlink[0]{}%
\providecommand \url  [0]{\begingroup\@sanitize@url \@url }%
\providecommand \@url [1]{\endgroup\@href {#1}{\urlprefix }}%
\providecommand \urlprefix  [0]{URL }%
\providecommand \Eprint [0]{\href }%
\providecommand \doibase [0]{http://dx.doi.org/}%
\providecommand \selectlanguage [0]{\@gobble}%
\providecommand \bibinfo  [0]{\@secondoftwo}%
\providecommand \bibfield  [0]{\@secondoftwo}%
\providecommand \translation [1]{[#1]}%
\providecommand \BibitemOpen [0]{}%
\providecommand \bibitemStop [0]{}%
\providecommand \bibitemNoStop [0]{.\EOS\space}%
\providecommand \EOS [0]{\spacefactor3000\relax}%
\providecommand \BibitemShut  [1]{\csname bibitem#1\endcsname}%
\let\auto@bib@innerbib\@empty
\bibitem [{\citenamefont {Dmitriev}\ \emph {et~al.}(2000)\citenamefont
  {Dmitriev}, \citenamefont {Krivnov},\ and\ \citenamefont
  {Ovchinnikov}}]{Dmitriev2000}%
  \BibitemOpen
  \bibfield  {author} {\bibinfo {author} {\bibfnamefont {D.}~\bibnamefont
  {Dmitriev}}, \bibinfo {author} {\bibfnamefont {V.}~\bibnamefont {Krivnov}}, \
  and\ \bibinfo {author} {\bibfnamefont {A.}~\bibnamefont {Ovchinnikov}},\
  }\href {\doibase 10.1007/s100510050110} {\bibfield  {journal} {\bibinfo
  {journal} {Eur Phys J B : Condensed Matter and Complex Systems}\ }\textbf
  {\bibinfo {volume} {14}},\ \bibinfo {pages} {91} (\bibinfo {year}
  {2000})}\BibitemShut {NoStop}%
\bibitem [{\citenamefont {Hida}\ \emph {et~al.}(2013)\citenamefont {Hida},
  \citenamefont {Takano},\ and\ \citenamefont {Suzuki}}]{Hida2013}%
  \BibitemOpen
  \bibfield  {author} {\bibinfo {author} {\bibfnamefont {K.}~\bibnamefont
  {Hida}}, \bibinfo {author} {\bibfnamefont {K.}~\bibnamefont {Takano}}, \ and\
  \bibinfo {author} {\bibfnamefont {H.}~\bibnamefont {Suzuki}},\ }\href
  {\doibase 10.7566/JPSJ.82.064703} {\bibfield  {journal} {\bibinfo  {journal}
  {J. Phys. Soc. Japan}\ }\textbf {\bibinfo {volume} {82}},\ \bibinfo {pages}
  {064703} (\bibinfo {year} {2013})}\BibitemShut {NoStop}%
\bibitem [{\citenamefont {Majumdar}\ and\ \citenamefont
  {Ghosh}(1969)}]{Majumder1969}%
  \BibitemOpen
  \bibfield  {author} {\bibinfo {author} {\bibfnamefont {C.~K.}\ \bibnamefont
  {Majumdar}}\ and\ \bibinfo {author} {\bibfnamefont {D.~K.}\ \bibnamefont
  {Ghosh}},\ }\href {\doibase 10.1063/1.1664979} {\bibfield  {journal}
  {\bibinfo  {journal} {J. Math. Phys.}\ }\textbf {\bibinfo {volume} {10}},\
  \bibinfo {pages} {1399} (\bibinfo {year} {1969})}\BibitemShut {NoStop}%
\bibitem [{\citenamefont {Chitra}\ \emph {et~al.}(1995)\citenamefont {Chitra},
  \citenamefont {Pati}, \citenamefont {Krishnamurthy}, \citenamefont {Sen},\
  and\ \citenamefont {Ramasesha}}]{Chitra1995}%
  \BibitemOpen
  \bibfield  {author} {\bibinfo {author} {\bibfnamefont {R.}~\bibnamefont
  {Chitra}}, \bibinfo {author} {\bibfnamefont {S.}~\bibnamefont {Pati}},
  \bibinfo {author} {\bibfnamefont {H.~R.}\ \bibnamefont {Krishnamurthy}},
  \bibinfo {author} {\bibfnamefont {D.}~\bibnamefont {Sen}}, \ and\ \bibinfo
  {author} {\bibfnamefont {S.}~\bibnamefont {Ramasesha}},\ }\href {\doibase
  10.1103/PhysRevB.52.6581} {\bibfield  {journal} {\bibinfo  {journal} {Phys.
  Rev. B}\ }\textbf {\bibinfo {volume} {52}},\ \bibinfo {pages} {6581}
  (\bibinfo {year} {1995})}\BibitemShut {NoStop}%
\bibitem [{\citenamefont {{Sriram Shastry}}\ and\ \citenamefont
  {Sutherland}(1981)}]{SRIRAMSHASTRY1981}%
  \BibitemOpen
  \bibfield  {author} {\bibinfo {author} {\bibfnamefont {B.}~\bibnamefont
  {{Sriram Shastry}}}\ and\ \bibinfo {author} {\bibfnamefont {B.}~\bibnamefont
  {Sutherland}},\ }\href {\doibase
  https://doi.org/10.1016/0378-4363(81)90838-X} {\bibfield  {journal} {\bibinfo
   {journal} {Physica B+C}\ }\textbf {\bibinfo {volume} {108}},\ \bibinfo
  {pages} {1069} (\bibinfo {year} {1981})}\BibitemShut {NoStop}%
\bibitem [{\citenamefont {Furukawa}\ \emph {et~al.}(2012)\citenamefont
  {Furukawa}, \citenamefont {Sato}, \citenamefont {Onoda},\ and\ \citenamefont
  {Furusaki}}]{Furukuwa2012}%
  \BibitemOpen
  \bibfield  {author} {\bibinfo {author} {\bibfnamefont {S.}~\bibnamefont
  {Furukawa}}, \bibinfo {author} {\bibfnamefont {M.}~\bibnamefont {Sato}},
  \bibinfo {author} {\bibfnamefont {S.}~\bibnamefont {Onoda}}, \ and\ \bibinfo
  {author} {\bibfnamefont {A.}~\bibnamefont {Furusaki}},\ }\href {\doibase
  10.1103/PhysRevB.86.094417} {\bibfield  {journal} {\bibinfo  {journal} {Phys.
  Rev. B}\ }\textbf {\bibinfo {volume} {86}},\ \bibinfo {pages} {094417}
  (\bibinfo {year} {2012})}\BibitemShut {NoStop}%
\bibitem [{\citenamefont {Sahoo}\ \emph {et~al.}(2020)\citenamefont {Sahoo},
  \citenamefont {Dey}, \citenamefont {Saha},\ and\ \citenamefont
  {Kumar}}]{Sahoo2020}%
  \BibitemOpen
  \bibfield  {author} {\bibinfo {author} {\bibfnamefont {S.}~\bibnamefont
  {Sahoo}}, \bibinfo {author} {\bibfnamefont {D.}~\bibnamefont {Dey}}, \bibinfo
  {author} {\bibfnamefont {S.~K.}\ \bibnamefont {Saha}}, \ and\ \bibinfo
  {author} {\bibfnamefont {M.}~\bibnamefont {Kumar}},\ }\href {\doibase
  10.1088/1361-648X/ab8663} {\bibfield  {journal} {\bibinfo  {journal} {J.
  Condens. Matter Phys.}\ }\textbf {\bibinfo {volume} {32}},\ \bibinfo {pages}
  {335601} (\bibinfo {year} {2020})}\BibitemShut {NoStop}%
\bibitem [{\citenamefont {Haldane}(1983)}]{Haldane1983}%
  \BibitemOpen
  \bibfield  {author} {\bibinfo {author} {\bibfnamefont {F.}~\bibnamefont
  {Haldane}},\ }\href {\doibase https://doi.org/10.1016/0375-9601(83)90631-X}
  {\bibfield  {journal} {\bibinfo  {journal} {Phys. Lett. A}\ }\textbf
  {\bibinfo {volume} {93}},\ \bibinfo {pages} {464} (\bibinfo {year}
  {1983})}\BibitemShut {NoStop}%
\bibitem [{\citenamefont {White}(1992)}]{White1992}%
  \BibitemOpen
  \bibfield  {author} {\bibinfo {author} {\bibfnamefont {S.~R.}\ \bibnamefont
  {White}},\ }\href {\doibase 10.1103/PhysRevLett.69.2863} {\bibfield
  {journal} {\bibinfo  {journal} {Phys. Rev. Lett.}\ }\textbf {\bibinfo
  {volume} {69}},\ \bibinfo {pages} {2863} (\bibinfo {year}
  {1992})}\BibitemShut {NoStop}%
\bibitem [{\citenamefont {Schollw\"ock}(2005)}]{Schollwock2005}%
  \BibitemOpen
  \bibfield  {author} {\bibinfo {author} {\bibfnamefont {U.}~\bibnamefont
  {Schollw\"ock}},\ }\href {\doibase 10.1103/RevModPhys.77.259} {\bibfield
  {journal} {\bibinfo  {journal} {Rev. Mod. Phys.}\ }\textbf {\bibinfo {volume}
  {77}},\ \bibinfo {pages} {259} (\bibinfo {year} {2005})}\BibitemShut
  {NoStop}%
\bibitem [{\citenamefont {Hallberg}(2006)}]{Hallberg2006}%
  \BibitemOpen
  \bibfield  {author} {\bibinfo {author} {\bibfnamefont {K.~A.}\ \bibnamefont
  {Hallberg}},\ }\href {\doibase 10.1080/00018730600766432} {\bibfield
  {journal} {\bibinfo  {journal} {Adv. Phys.}\ }\textbf {\bibinfo {volume}
  {55}},\ \bibinfo {pages} {477} (\bibinfo {year} {2006})}\BibitemShut
  {NoStop}%
\bibitem [{\citenamefont {Saha}\ \emph {et~al.}(2019)\citenamefont {Saha},
  \citenamefont {Dey}, \citenamefont {Kumar},\ and\ \citenamefont
  {Soos}}]{Sudip2019}%
  \BibitemOpen
  \bibfield  {author} {\bibinfo {author} {\bibfnamefont {S.~K.}\ \bibnamefont
  {Saha}}, \bibinfo {author} {\bibfnamefont {D.}~\bibnamefont {Dey}}, \bibinfo
  {author} {\bibfnamefont {M.}~\bibnamefont {Kumar}}, \ and\ \bibinfo {author}
  {\bibfnamefont {Z.~G.}\ \bibnamefont {Soos}},\ }\href {\doibase
  10.1103/PhysRevB.99.195144} {\bibfield  {journal} {\bibinfo  {journal} {Phys.
  Rev. B}\ }\textbf {\bibinfo {volume} {99}},\ \bibinfo {pages} {195144}
  (\bibinfo {year} {2019})}\BibitemShut {NoStop}%
\bibitem [{\citenamefont {Parvej}\ and\ \citenamefont
  {Kumar}(2016)}]{Parvej2016}%
  \BibitemOpen
  \bibfield  {author} {\bibinfo {author} {\bibfnamefont {A.}~\bibnamefont
  {Parvej}}\ and\ \bibinfo {author} {\bibfnamefont {M.}~\bibnamefont {Kumar}},\
  }\href {\doibase https://doi.org/10.1016/j.jmmm.2015.10.017} {\bibfield
  {journal} {\bibinfo  {journal} {J. Magn. Magn. Mater.}\ }\textbf {\bibinfo
  {volume} {401}},\ \bibinfo {pages} {96} (\bibinfo {year} {2016})}\BibitemShut
  {NoStop}%
\bibitem [{\citenamefont {Onishi}(2015)}]{Onishi2015}%
  \BibitemOpen
  \bibfield  {author} {\bibinfo {author} {\bibfnamefont {H.}~\bibnamefont
  {Onishi}},\ }\href {\doibase 10.7566/JPSJ.84.083702} {\bibfield  {journal}
  {\bibinfo  {journal} {J. Phys. Soc. Japan}\ }\textbf {\bibinfo {volume}
  {84}},\ \bibinfo {pages} {083702} (\bibinfo {year} {2015})}\BibitemShut
  {NoStop}%
\bibitem [{\citenamefont {Parvej}\ and\ \citenamefont
  {Kumar}(2017)}]{Parvej2017}%
  \BibitemOpen
  \bibfield  {author} {\bibinfo {author} {\bibfnamefont {A.}~\bibnamefont
  {Parvej}}\ and\ \bibinfo {author} {\bibfnamefont {M.}~\bibnamefont {Kumar}},\
  }\href {\doibase 10.1103/PhysRevB.96.054413} {\bibfield  {journal} {\bibinfo
  {journal} {Phys. Rev. B}\ }\textbf {\bibinfo {volume} {96}},\ \bibinfo
  {pages} {054413} (\bibinfo {year} {2017})}\BibitemShut {NoStop}%
\bibitem [{\citenamefont {Sato}\ \emph {et~al.}(2011)\citenamefont {Sato},
  \citenamefont {Hikihara},\ and\ \citenamefont {Momoi}}]{Sato2011}%
  \BibitemOpen
  \bibfield  {author} {\bibinfo {author} {\bibfnamefont {M.}~\bibnamefont
  {Sato}}, \bibinfo {author} {\bibfnamefont {T.}~\bibnamefont {Hikihara}}, \
  and\ \bibinfo {author} {\bibfnamefont {T.}~\bibnamefont {Momoi}},\ }\href
  {\doibase 10.1088/1742-6596/320/1/012014} {\bibfield  {journal} {\bibinfo
  {journal} {J. Phys. Conf. Ser}\ }\textbf {\bibinfo {volume} {320}},\ \bibinfo
  {pages} {012014} (\bibinfo {year} {2011})}\BibitemShut {NoStop}%
\bibitem [{\citenamefont {White}\ and\ \citenamefont {Huse}(1993)}]{White1993}%
  \BibitemOpen
  \bibfield  {author} {\bibinfo {author} {\bibfnamefont {S.~R.}\ \bibnamefont
  {White}}\ and\ \bibinfo {author} {\bibfnamefont {D.~A.}\ \bibnamefont
  {Huse}},\ }\href {\doibase 10.1103/PhysRevB.48.3844} {\bibfield  {journal}
  {\bibinfo  {journal} {Phys. Rev. B}\ }\textbf {\bibinfo {volume} {48}},\
  \bibinfo {pages} {3844} (\bibinfo {year} {1993})}\BibitemShut {NoStop}%
\bibitem [{\citenamefont {Hida}(1992)}]{Hida1992}%
  \BibitemOpen
  \bibfield  {author} {\bibinfo {author} {\bibfnamefont {K.}~\bibnamefont
  {Hida}},\ }\href {\doibase 10.1103/PhysRevB.45.2207} {\bibfield  {journal}
  {\bibinfo  {journal} {Phys. Rev. B}\ }\textbf {\bibinfo {volume} {45}},\
  \bibinfo {pages} {2207} (\bibinfo {year} {1992})}\BibitemShut {NoStop}%
\bibitem [{\citenamefont {Allen}\ and\ \citenamefont
  {S\'en\'echal}(1997)}]{Allen1997}%
  \BibitemOpen
  \bibfield  {author} {\bibinfo {author} {\bibfnamefont {D.}~\bibnamefont
  {Allen}}\ and\ \bibinfo {author} {\bibfnamefont {D.}~\bibnamefont
  {S\'en\'echal}},\ }\href {\doibase 10.1103/PhysRevB.55.299} {\bibfield
  {journal} {\bibinfo  {journal} {Phys. Rev. B}\ }\textbf {\bibinfo {volume}
  {55}},\ \bibinfo {pages} {299} (\bibinfo {year} {1997})}\BibitemShut
  {NoStop}%
\bibitem [{\citenamefont {{Mattis}}\ and\ \citenamefont
  {{Lieb}}(1965)}]{Mattis1965}%
  \BibitemOpen
  \bibfield  {author} {\bibinfo {author} {\bibfnamefont {D.~C.}\ \bibnamefont
  {{Mattis}}}\ and\ \bibinfo {author} {\bibfnamefont {E.~H.}\ \bibnamefont
  {{Lieb}}},\ }\href {\doibase 10.1063/1.1704281} {\bibfield  {journal}
  {\bibinfo  {journal} {J. Math. Phys.}\ }\textbf {\bibinfo {volume} {6}},\
  \bibinfo {pages} {304} (\bibinfo {year} {1965})}\BibitemShut {NoStop}%
\bibitem [{\citenamefont {Okamoto}\ and\ \citenamefont
  {Nomura}(1992)}]{nomura1992}%
  \BibitemOpen
  \bibfield  {author} {\bibinfo {author} {\bibfnamefont {K.}~\bibnamefont
  {Okamoto}}\ and\ \bibinfo {author} {\bibfnamefont {K.}~\bibnamefont
  {Nomura}},\ }\href {\doibase https://doi.org/10.1016/0375-9601(92)90823-5}
  {\bibfield  {journal} {\bibinfo  {journal} {Phys. Lett. A}\ }\textbf
  {\bibinfo {volume} {169}},\ \bibinfo {pages} {433} (\bibinfo {year}
  {1992})}\BibitemShut {NoStop}%
\bibitem [{\citenamefont {Soos}\ \emph {et~al.}(2016)\citenamefont {Soos},
  \citenamefont {Parvej},\ and\ \citenamefont {Kumar}}]{soos2016}%
  \BibitemOpen
  \bibfield  {author} {\bibinfo {author} {\bibfnamefont {Z.~G.}\ \bibnamefont
  {Soos}}, \bibinfo {author} {\bibfnamefont {A.}~\bibnamefont {Parvej}}, \ and\
  \bibinfo {author} {\bibfnamefont {M.}~\bibnamefont {Kumar}},\ }\href
  {\doibase 10.1088/0953-8984/28/17/175603} {\bibfield  {journal} {\bibinfo
  {journal} {J. Phys. Condens. Matter}\ }\textbf {\bibinfo {volume} {28}},\
  \bibinfo {pages} {175603} (\bibinfo {year} {2016})}\BibitemShut {NoStop}%
\end{thebibliography}%
\end{document}